\documentclass[pdflatex,sn-mathphys,Numbered]{sn-jnl}

\usepackage{graphicx}%
\usepackage{multirow}%
\usepackage{amsmath,amssymb,amsfonts}%
\usepackage{amsthm}%
\usepackage{mathrsfs}%
\usepackage[title]{appendix}%
\usepackage{xcolor}%
\usepackage{textcomp}%
\usepackage{manyfoot}%
\usepackage{booktabs}%
\usepackage{algorithm}%
\usepackage{algorithmicx}%
\usepackage{algpseudocode}%
\usepackage{listings}%
\usepackage{titlesec}
\usepackage{graphicx}
\usepackage{hyphenat}
\usepackage{pdfpages}
\usepackage{float}
\usepackage{color,soul}
\usepackage{amsmath}
\usepackage{amssymb}
\usepackage{physics}
\usepackage{textcomp, gensymb}
\usepackage{gensymb} 
\usepackage{hyperref}
\usepackage{comment}
\hypersetup{
    colorlinks=true,
    linkcolor=blue,
    filecolor=magenta,      
    urlcolor=cyan,
    pdftitle={Overleaf Example},
    pdfpagemode=FullScreen,
    }
\usepackage[section]{placeins}
\usepackage{latexsym}

\begin{document}

\title[Article Title]{Depth-dependent warming of the Gulf of Eilat (Aqaba)}


\author*[1]{\fnm{Sounav} \sur{Sengupta}}\email{sounav@post.bgu.ac.il}

\author[2]{\fnm{Hezi} \sur{Gildor}}\email{hezi.gildor@mail.huji.ac.il}

\author[1]{\fnm{Yosef} \sur{Ashkenazy}}\email{ashkena@bgu.ac.il}

\affil*[1]{\orgdiv{Department of Environmental Physics, BIDR}, \orgname{Ben-Gurion University of the Negev}, \country{Israel}}

\affil[2]{\orgdiv{The Institute of Earth Sciences}, \orgname{The Hebrew University of Jerusalem}, \country{Israel}}


\abstract{The Gulf of Eilat (Gulf of Aqaba) is a semi-enclosed basin situated at the northern end of the Red Sea, renowned for its exceptional marine ecosystem. To evaluate the response of the Gulf to climate variations, we analyzed various factors including temperature down to 700 m, surface air temperature, and heat fluxes. We find that the sea temperature is rising at all depths despite inconclusive trends in local atmospheric variables, including the surface air temperature. The Gulf's sea surface temperature warms at a rate of a few hundredths of a degree Celsius per year, which is comparable to the warming of the global sea surface temperature and the Mediterranean Sea. The increase in sea warming is linked to fewer winter deep mixing events that used to occur more frequently in the past. Based on the analysis of the ocean-atmosphere heat fluxes, we conclude that the lateral advection of heat from the southern part of the Gulf likely leads to an increase in water temperature in the northern part of the Gulf. Our findings suggest that local ocean warming is not necessarily associated with local processes, but rather with the warming of other remote locations. 
}




\maketitle

\begin{figure}[ht]
    \centering
    \advance\leftskip-0.0cm
    \includegraphics[scale=0.3]{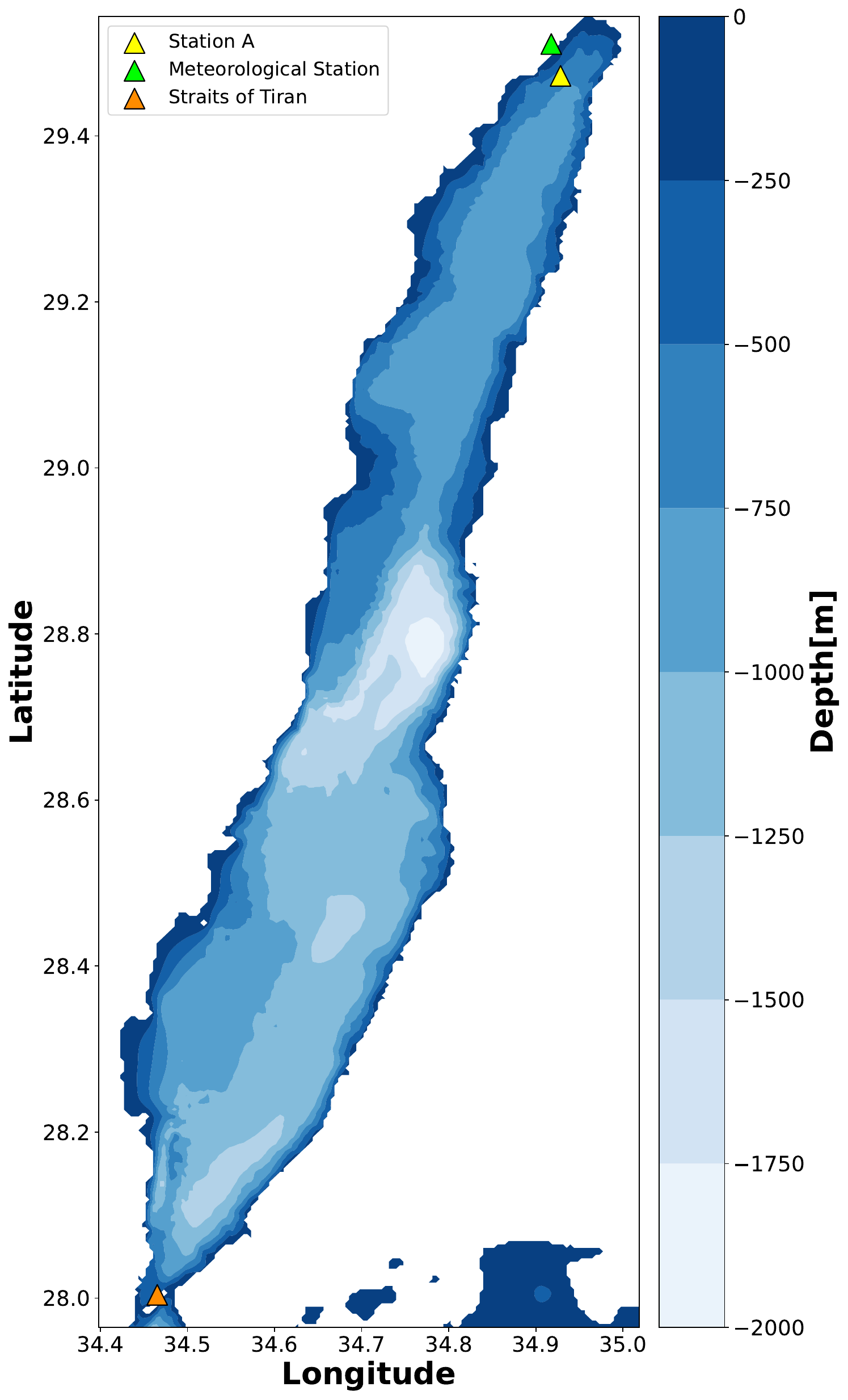}
\caption{A depth contour map of the Gulf of Eilat and the locations of the oceanic Station A (yellow triangle), the meteorological station at the Inter-University Institute (green triangle), and the Straits of Tiran (orange triangle at the bottom).}
\label{fig:1}     
\end{figure}

\section{Introduction}
\label{intro}

The warming of sea and air temperatures in different parts of the world is well established. According to the recent Intergovernmental Panel on Climate Change (IPCC) \cite{ipcc2023} the Earth's oceans continue to warm due to human activity. It was reported that the ocean temperature has increased over the past few decades, absorbing more than 90\% of the excess heat in the Earth's system during this period. This increase in heat has caused the ocean to expand and contributed to $\sim$43\% of observed global mean sea level rise between 1970 and 2015. Newer estimates of ocean heat uptake in the top 2000 m between 1993 and 2017 range from 9.2$\pm$2.3 ZJ/yr to 12.1$\pm$3.1 ZJ/yr \cite{ipcc2014}; 1 zettajoule (ZJ) equals $10^{21}$ J. The updated estimates for ocean heat uptake in the upper ocean (0–700 meters) indicate warming with more recent periods experiencing faster rates of warming. In addition, the rate of heat uptake for deeper layers has been higher in the last three decades \cite{ipcc2014}. 

A warming trend in the global subsurface ocean has been identified, separate from natural variations. This warming pattern consists of three main components: long-term warming, El Niño/Southern Oscillation-related (ENSO) warming \cite{nisha2024trend}, and Atlantic Multidecadal Oscillation-related warming \cite{xie2020ocean,yang2020global}. Long-term warming accounts for 78\% of the global temperature variance down to 300 m depth \cite{Yasunaka2012}. In addition, the rate of increase in global average Sea Surface Temperature (SST) is around 0.033\degree{C}/yr, based on data analyzed from 1982 to 2023 \cite{mediterranean}. During this same period, the Mediterranean Sea has experienced an increase of around 0.036\degree{C}/yr \cite{mediterranean,Bulgin2020}. 

Nisha et al. \cite{nisha2024trend} examine a long-term trend and interannual variability of heat content down to 300 of the Arabian Sea (2000-2017), attributing spring/summer warming to subsurface heat accumulation, and fall/winter to mixed layer heat. El Niño/Southern Oscillation (ENSO) and Indian Ocean Dipole (IOD) drive the interannual variability below the mixed layer, while advection and air-sea fluxes impact different seasons, emphasizing the role of these dynamics in marine heatwaves, cyclones, and regional climate predictions.

The Gulf of Eilat (hereafter the ``Gulf'') hosts a diverse ecosystem, and it was suggested that the Gulf may be the last refuge for coral reefs \cite{fine2013coral}. Other studies addressed the ocean warming's impact on the Gulf's ecosystem  \cite{Abir2022,Miller2017,Labiosa:2003}. Here we investigate the presence of discernible trends within the temperature profiles of the water column within the Gulf. Our analysis reveals a rising temperature trend throughout the water column, a trend seemingly disconnected from immediate local atmospheric conditions. Instead, it appears to be intricately linked to the advection of water from southern latitudes, implying a plausible scenario of warming occurring within the Red Sea \cite{Wolf-Vecht:1992}. This horizontal advection of water from the Red Sea is confined by the shallow sills and an upper warm layer of water with a weak vertical stratification \cite{Reiss1984,Paldor:1979,Wolf-Vecht:1992,Labiosa:2003,Biton-Gildor}. Precisely estimating the heat and water exchange between the atmosphere and the ocean is crucial for gaining insights into thermoregulation mechanisms and for developing accurate models of ocean dynamics \cite{Abir2023}. Hence, a long-term analysis of sea and air temperature, like the analysis performed in this study, can facilitate a deeper understanding of the climate system and its effect on the marine ecosystem.

The distinctive statistical methodology employed herein underscores the robustness of our conclusions regarding the substantive warming trends observed over the years of sampled data. Below we first provide a brief introduction to the Gulf (Sec.~\ref{study-area}). Afterward, we describe the oceanic (Sec.~\ref{sec:1.1}) and meteorological (Sec.~\ref{sec:1.2}) datasets, and the methodology employed (Sec.~\ref{sec:1.3}). We then present the results that are based on oceanic data (Sec.~\ref{sec:3.1}) and meteorological data (Sec.~\ref{sec:3.2}), including the estimation of surface heat fluxes in the Gulf (Sec.~\ref{sec:3.3}). We discuss and conclude the study in Sec.~\ref{sec:summary}.

\section{The study area: The Gulf of Eilat (Aqaba)}
\label{study-area}

The Gulf is situated at the northern tip of the Red Sea and is surrounded by the Sinai and the Arabian peninsula (Fig.~\ref{fig:1}). It is an elongated semi-enclosed basin that is approximately 180 km long, 5-25 km wide, and has a maximum depth of 1820 m. The Gulf is connected to the Red Sea through the Straits of Tiran (orange triangle in Fig. \ref{fig:1}), which have a maximum depth of 245 m \cite{Reiss1984,Murray:1984}. The Gulf abodes a diverse ecosystem of coral reefs and their associated fisheries, that are under the influence of significant anthropogenic stress 
\cite{Baharatan2010,Dishon2012,Miller2017,genin2020rapid}. 

\begin{figure}[t]
    \centering
    \includegraphics[width=\textwidth]{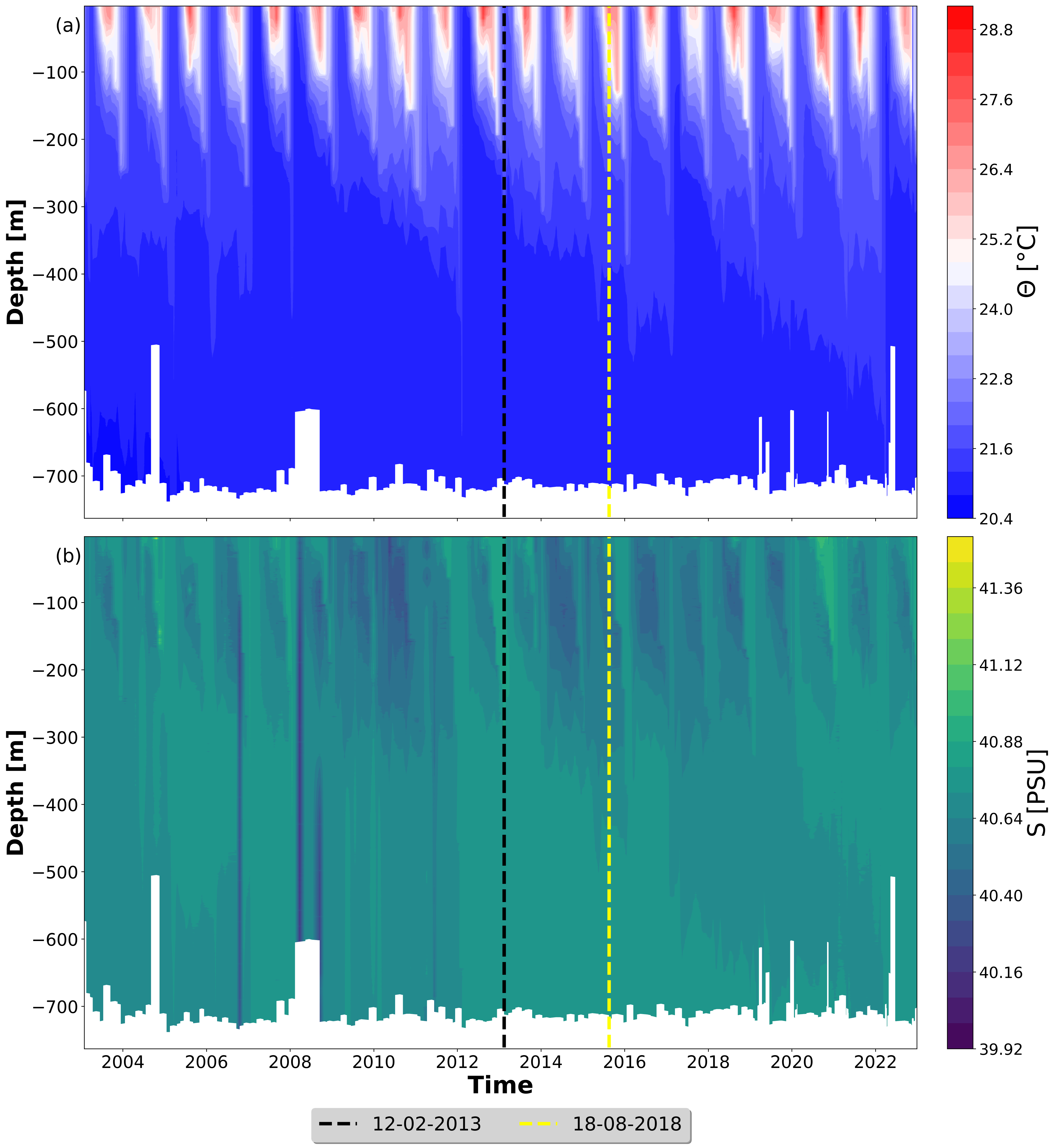}
    \caption{(a) Water potential temperature (in \degree{C}) and (b) salinity (in PSU) as a function of time and depth. The black and yellow vertical dashed lines indicate two specific dates whose potential temperature and salinity profiles are plotted in Fig.~\ref{fig:3}. The above measurements were collected every month at Station A in the northern part of the Gulf of Eilat (indicated by the yellow triangle in Fig.~\ref{fig:1}), between January 2003 and December 2022.}
    \label{fig:2}     
\end{figure}

The water column's temperature profile in the northern Gulf exhibits pronounced seasonal variations. During the summer and autumn seasons, from June to October, the upper mixed layer of the water column extends down to a depth of $\sim$40 m, with SST of around 27\degree{C}. Below this mixed layer lies a stratified thermocline, extending to a depth of approximately 200 m \cite{Paldor:1979}. Progressing further downward is a deep, nearly uniform layer that extends to the seafloor (reaching depths of over 700 m at the specified measurement location) and maintains a temperature of just under 21\degree{C}; see Fig.~\ref{fig:2}.

\begin{figure}[t]
    \centering
    \advance\leftskip-1.75cm
    \includegraphics[width=\textwidth]{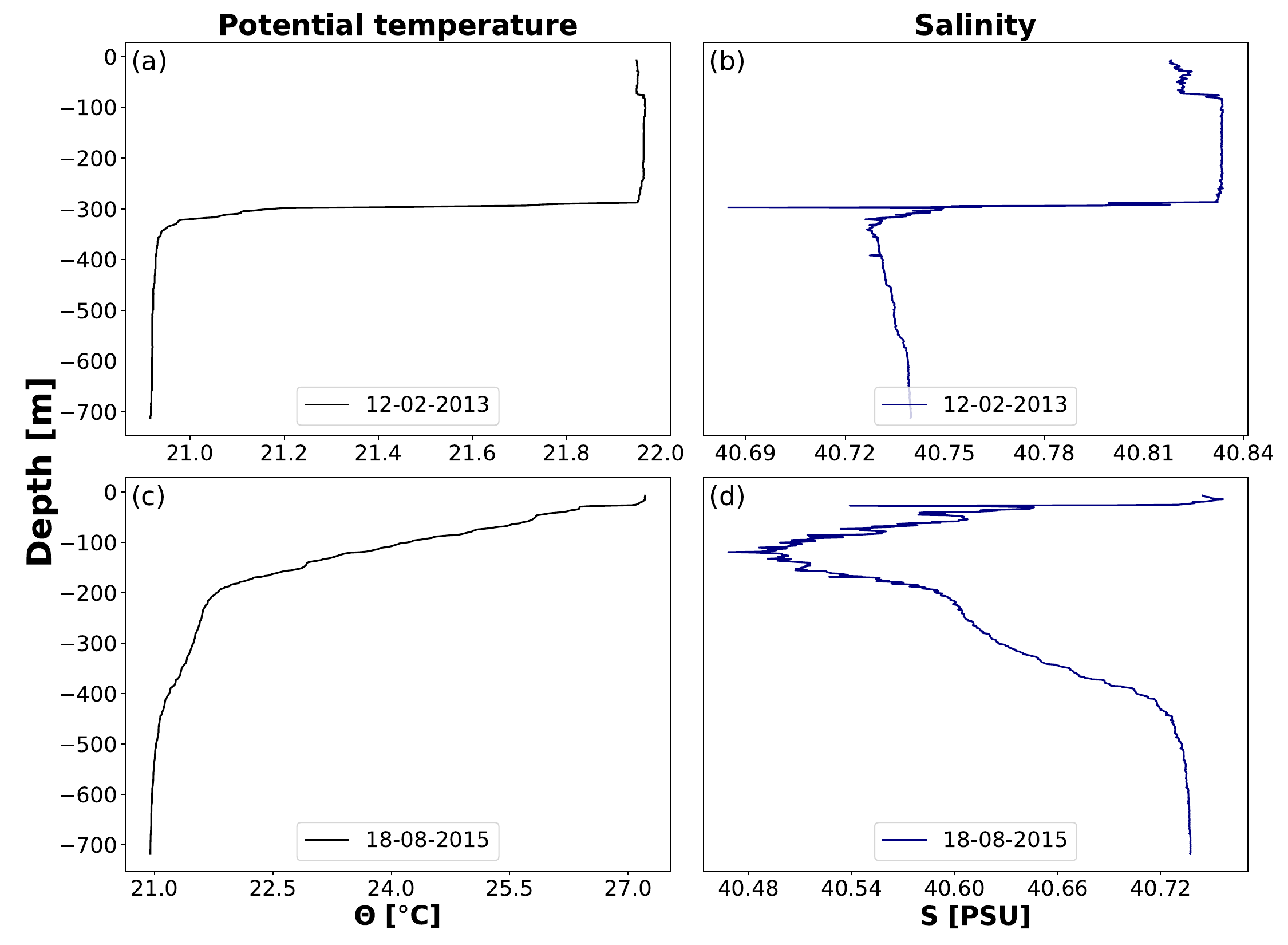}
    \caption{Ocean potential temperature (a,c) and salinity (b,d) depth profiles at (a,b) a typical winter month (12-02-2013) and (c,d) a typical summer month (18-08-2015). These dates are marked by the black and yellow vertical dashed lines in Fig.~\ref{fig:2}.}
    \label{fig:3}    
\end{figure}

As the autumn season unfolds between October and December, the SST decreases due to atmospheric cooling. This cooling triggers, through vertical convection, a deepening of the mixed layer. Subsequently, during the winter season, from January to March, this vertical mixing process intensifies, causing the thermocline to largely dissipate. Consequently, a weak temperature contrast of less than 1\degree{C} prevails between surface water and deep water in most years; see Fig. \ref{fig:3}a. However, particularly cold years witness the complete erosion of the thermocline due to extensive deep mixing, resulting in the formation of a singular deep layer with uniform temperature and salinity \cite{NathanPaldor,Paldor:1979}; see Fig.~\ref{fig:2}. Such complete mixing events were observed on the following dates: 12-02-2007, 12-02-2008, 13-03-2012, and 23-03-2022, indicating only one complete mixing event in the last 10 years. 

The changes in the stratification of the Gulf across the seasons have significant impacts on both the Gulf's water dynamics and ecosystem. Seasonal changes influence the exchange flow with the Red Sea and the stratification in the Gulf \cite{biton2011coupling}. The change in stratification also affects the magnitude of semi-diurnal tidal currents \cite{monismith2004tides,Carlson2012}. In winter, when the surface water is sufficiently cold and dense, deep mixing occurs, which permeates the deep waters and mixes nutrients into the photic zone. This nutrient enrichment triggers blooms of phytoplankton and zooplankton that peak at the beginning of the restratification phase, towards the end of the winter \cite{Genin:1995,Berman2022,Berman2023}.

\section{Data and Methods}

\subsection{Ocean Data}
\label{sec:1.1}

The ocean data consists of 270 standard CTD casts measured in ``Station A'' (29\degree 28' N, 34\degree 55' E, indicated by the yellow triangle in Fig.~\ref{fig:1}). The data was collected on monthly cruises as part of the Israeli national ocean monitoring program. Each cast includes measurements of pressure, in-situ temperature, conductivity, photosynthetic active radiation (PAR), and fluorescence. To study the deep water temperature temporal variations, we consider casts that extend to depths below 500 m. We binned the data into a one-meter resolution, to allow consistent statistical analysis. The ocean data (potential temperature and salinity) we analyzed below is depicted in Fig.~\ref{fig:2}, typically 1 cast per month, from January 2003 to December 2022 (20 years). The maximum depth of these measurements is $\sim$750 m (see Fig.~\ref{fig:1}). We note that casts have been conducted at different phases of the tides which stimulate internal waves which can shift up and down the isotherms by a few tens of meters \cite{CARLSON20141}. This is expected to increase the variability of the temperature at the thermocline, and hence to increase the uncertainty of our results. Despite this, we find a significant warming along the entire water column.

\subsection{Meteorological Data}
\label{sec:1.2}
The meteorological data has been recorded at the end of the dock of the Israeli Inter-University Institute (IUI) for Marine Sciences, Eilat (29\degree 30.211’ E, 34\degree 55.068’ N, indicated by the green triangle in Fig.~\ref{fig:1}), where a comprehensive set of environmental parameters are being measured. The main mast, situated 10 m above sea level at low tide, serves as the reference point for recording wind speed, wind direction, air temperature, and relative humidity. Barometric pressure is measured 5 m above sea level, and adjustments are made to ensure sea level calibration. For capturing solar radiation data, a sensor is fixed on a cross-arm extending 0.7 m southward from the main mast, positioned 6 m above sea level. 

The atmospheric data we analyzed span the period from the beginning of January 2007 to the end of December 2022 (16 years). The collected data included a 10-minute average of the different variables including (10 m) surface air temperature (SAT), SST (measured at a 2~m depth), sea level pressure, solar radiation, 10~m wind speed and direction, relative and absolute humidity, photosynthetic active radiation (PAR), and UV radiation. We constructed the daily and monthly averages of the data which have been used to calculate the heat fluxes and their climatological seasonal cycle.

\subsection{Methods}
\label{sec:1.3}

\subsubsection{Ocean data analysis}
\label{sec:1.3.1}

In our study, we aim to investigate possible long-term temporal trends (i.e., beyond the seasonal cycle) in various environmental variables (including the SAT and various heat fluxes discussed below), with a particular focus on water temperature. To accomplish this, we employed the method of least square linear regression on data points (time series) binned at different depths. It is essential to emphasize that we do not assume the trends are linear; rather, we used linear regression solely to quantify the general increase or decrease of the different variables over the duration of the measurements. 

To assess the significance of these trends, we implemented the following statistical approach. We conducted random shuffling of the time series for each variable (at each depth), a process we repeated many times (typically 1000 times). By shuffling the time series, we effectively destroy the temporal order and thereby render it random \cite{Wood2018,Lancaster2018}. Consequently, any genuine temporal patterns (including trends) that might have existed in the original data were disrupted in the shuffling procedure. We then performed linear regression on each of the shuffled time series to calculate the slopes of their respective trends. The mean of these slopes should be around zero as there are no correlations in the shuffled time series. By constructing a Probability Density Function (PDF) of the slopes obtained from the shuffled time series, we generated an essential tool for evaluating the significance of the trend slopes observed in the original data. If the actual slope of the trend in the original time series lies well outside the confidence interval of the PDF generated from the shuffled slopes, it suggests that the observed trend is statistically significant, up to the confidence level. In other words, the probability of obtaining such a slope by random chance alone is exceedingly low, providing evidence for a meaningful temporal trend in the variable under consideration \cite{bootstrap}. This methodology ensures that our conclusions regarding the significance of the observed trends are robust. The above procedure can be applied both for oceanic and meteorological data.

Both the different atmospheric and oceanic variables exhibit seasonal cycles that may make it difficult to estimate the temporal trends superimposed on the seasonal cycle. In light of the above, we have performed the estimations of the linear slope based on: (a) the original time series that includes the seasonal cycles, (b) the time series of a specific month (or season) over the years (e.g., considering the month of February of all years), and (c) on the anomalies around the seasonal cycle. It is clear that the uncertainty is the largest for (a); yet, for deep ocean measurements, the seasonal cycle is hardly present. We elaborate on the above below. 

\subsubsection{Meteorological Data Analysis}
\label{sec:1.3.2}

We conducted a comprehensive analysis of meteorological data collected at 10-minute intervals over 16 years (2007-2022). Our primary objective was to investigate long-term climate trends and understand the heat fluxes associated with the observed meteorological conditions. To achieve this, we computed the daily, monthly, and yearly averages for each of the atmospheric variables. This step allowed us to capture seasonal and annual variations in the meteorological parameters over the 16-year study period. Next, we employed bulk formulas to calculate the heat fluxes based on the derived averages.

\subsubsection{Heat Flux Analysis}
\label{1.3.3}

The Gulf is connected to the Red Sea through the Straits of Tiran and the water exchange through the straits affects the circulation and the structure of the water column in the Gulf \cite{Biton2023,berman2003seasonality}. The energy balance in the Gulf is affected both by the local meteorological conditions and the water properties that are influenced by the water exchange through the straits of Tiran. The principle of energy conservation helps to understand the thermoregulation within a marine setting. The net heat flux, $Q$, of a water column can be formulated as follows \cite{NathanPaldor,Abir2023}
\begin{equation}
    \begin{split}
        Q & =\mathrm{SW}+\mathrm{LW}+\mathrm{LH}+\mathrm{SH}+\mathrm{ADV} \\
        & =\mathrm{ASHF}+\mathrm{ADV},
    \end{split}
\end{equation}
where $\mathrm{SW}$ indicates the net shortwave (or solar) radiation, $\mathrm{LW}$ the net longwave radiation, LH the latent heat flux, $\mathrm{SH}$ the sensible heat flux, and $\mathrm{ADV}$ the heat transported by advection from the adjacent ocean. The sum of these components quantifies the net air-sea heat flux (ASHF) through the surface where positive values indicate that the water is losing heat (i.e., cooling down). While the shortwave radiation, $\mathrm{SW}$, is directly measured, the other fluxes can be estimated using standard bulk formulas specified below \cite{Taylor:2000}.

The latent and sensible heat fluxes can be estimated \cite{Fairall,Gill-1982:atmosphere,Smith:1988,Kondo:1975} based on the wind speed and differences in humidity and temperature between the sea surface and the air surface:
\begin{equation}
    \begin{split}
        \mathrm{LH}&=C_{e} \rho L\left(q_{s}-q_{a}\right) W, \\
        \mathrm{SH}&=C_{h} \rho C_{p}\left(T_{s}-T_{a}\right) W,
    \end{split}
    \label{eq:LHSH}
\end{equation}
where $L$ denotes the latent heat constant of vaporization ($2.5 \times 10^{6} ~ \mathrm{J}\, \mathrm{kg}^{-1}$), $\rho$ is the air density ($1.2 \mathrm{~kg}\, \mathrm{m}^{-3}$), $q_{s}$ and $q_{a}$ are the saturated specific humidity calculated using the SST ($T_s$) and surface air temperature ($T_a$), $W$ is the wind speed, $C_{e}$ and $C_{h}$ are the moisture and heat transfer coefficients, and $C_p$ is the specific heat constant of air. 

The longwave radiation has been parameterized previously based on the calculation of water vapor pressure and the atmospheric emissivity \cite{fung}. There are at least eight formulas that are used to calculate the longwave radiation, some of them incorporate a cloudiness factor for a more accurate estimation \cite{Josey:1997,Dickey:1994}. Here we use the formula of Ref. \cite{Berliand:1952} to estimate the longwave radiation; we chose a cloudiness factor of $c=1$ since the Gulf has fairly clear sky conditions throughout the year:
\begin{equation}
    \mathrm{LW} =  -\epsilon \sigma T_{a}^4 (0.39 - 0.05 \sqrt{e_a})c - 4 \epsilon \sigma T_{a}^3 (T_{s} - T_{a}),
    \label{eq:LW}
\end{equation}
where $\epsilon(=0.985)$ and $\sigma(=5.6697\times10^{-8}$ ${\rm Wm}^{-2}{\rm K}^{-4})$ are the emissivity and Stefan-Boltzmann constants respectively. $T_s$ and $T_a$ are the SST and surface air temperature in K and $e_a$ is the water vapor pressure at a (2~m) standard height.

\section{Results}

\subsection{Temporal trends and their significance} 
\label{sec:3.1}

Fig.~\ref{fig:2} presents the temperature and salinity at Station A of the Gulf as a function of time and depth. The temperature profile shown in Fig.~\ref{fig:2}a exhibits a clear seasonal cycle with deep mixing of several hundred meters during the winter months (February and March) and a relatively shallow mixed layer of tens of meters during the summer (July and August) \cite{CARLSON20141,Sofianos:2002}. The difference in SST between the summer and winter is about 7\degree{C} and the mixed layer is deepening towards the winter months when the SST becomes colder. Every few years, the ocean experiences a complete mixing down to the bottom, while every year there is mixing to at least a few hundred meters; see Fig.~\ref{fig:3}a. Below we show that deep mixing events become less frequent, leading to the warming of the entire water column in the Gulf.

The clear seasonal cycle observed in the water temperature is less apparent in the salinity field (Fig.~\ref{fig:2}b) \cite{Berman:2003} [We note that the salinity data till 2009 may contain some inaccuracies which may affect the salinity results reported below]. Moreover, salinity profiles are not monotonic, and there is a layer with a salinity minimum during the summer (e.g., Fig.~\ref{fig:3}d), due to the minimum in the salinity of the water entering through the Tiran Straits \cite{Biton-Gildor,biton2011coupling}. The salinity variations in the Gulf span 40.3 PSU to 40.8 PSU \cite{Genin:1995} while the temperature variations span 21\degree{C} to 28\degree{C}; simple analysis using a linear equation of state indicates that the density in the Gulf is temperature dominated \cite{Biton-Gildor}. This is further justified as follows: we estimated, based on the GSW package (\url{https://www.teos-10.org/pubs/gsw/html/gsw_contents.html}), the thermal and haline expansion coefficients, $\alpha$ and $\beta$, for typical temperature (25\degree{C}) and salinity (41 PSU) values as $\alpha=3.1\times 10^{-4}$~\degree{C}$^{-1}$ and $\beta=7.2\times 10^{-4}$~PSU$^{-1}$. Thus, $\alpha\Delta T/\beta\Delta S\gtrsim 3$, indicating that the role of temperature variations on the density is more than 3 times larger than the contribution of the salinity. Following the above, and since the observed trend in the water is a warming trend, we mainly focus on the temperature variations.

To have a clear view of the structure of the water column, we show in Fig.~\ref{fig:3} the potential temperature and salinity depth profiles in typical winter (12-02-2013) and summer (18-08-2015) dates; these are marked by the vertical black and yellow dashed lines in Fig.~\ref{fig:2}. The different water column layers are noticeable in the summer month temperature profile (Fig.~\ref{fig:3}c). The deep mixing of several hundred ($\sim$300) meters is shown in (Fig.~\ref{fig:3}a,b). 

\begin{figure}[t]
    \centering
    \includegraphics[width=\textwidth]{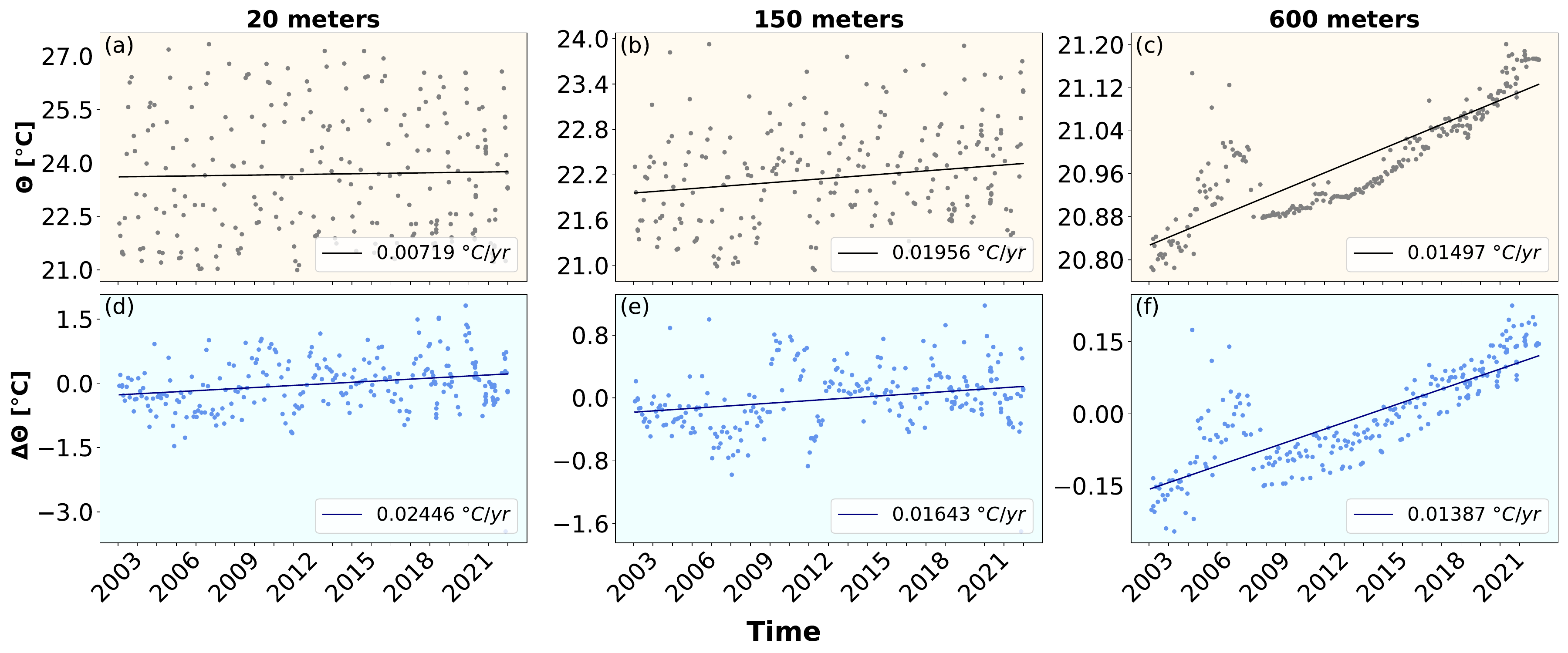}
    \caption{Water potential temperature as a function of time at depths of (a,d) 20 m, (b,e) 150 m, and (c,f) 600 m. The solid black lines indicate the linear regression where the slope is indicated in the lower right corner of each panel. The upper panels depict the original time series while the lower panels depict the anomaly time series after removing the seasonal cycle. Note the different scales of the y-axis in the different panels.}
    \label{fig:4}     
\end{figure}

We next analyze the temporal trends in oceanic temperature across the duration of the dataset. Fig.~\ref{fig:4} depicts the water potential temperature and potential temperature anomalies as a function of time for distinct depth levels: shallow depths (20 m, Fig.~\ref{fig:4}a,d), intermediate depths (150 m, Fig.~\ref{fig:4}b,e), and deep ocean (600 m, Fig.~\ref{fig:4}c,f). Notably, the upper panels of Fig.~\ref{fig:4} are augmented by linear regression lines, enabling the identification of nuanced trends within the dataset, often concealed by the dominant seasonal oscillations. It becomes evident that the temperature trends within the shallow and intermediate waters are considerably influenced by the seasonal cycle. However, a distinctive narrative unfolds within the deep ocean regions, where the influence of seasonal variability is notably diminished. Here, a discernible and statistically significant warming trend emerges. This trend is characterized by an annual temperature increase of approximately 0.015 \degree{C}\,yr$^{-1}$.

To filter out the effect of the seasonal cycle, we have estimated the temporal trends in the ocean temperature after removing the seasonal cycle from the time series shown in Fig.~\ref{fig:4}a-c. The results are plotted in Fig.~\ref{fig:4}d-f, indicating clearer increasing trends in ocean temperature. We verify the significance of these trends below. The increase in water temperature range from $\sim$0.015 \degree{C}\,yr$^{-1}$ at deep water, to $\sim$0.014 \degree{C}\,yr$^{-1}$ at depth of 150\,m, to $\sim$0.025 \degree{C}\,yr$^{-1}$ at shallow water of 20\,m. 

To estimate the significance of the temperature trends (like the ones shown in Fig. \ref{fig:4}), we conducted the hypothesis test described above in the Sec.~\ref{sec:1.3.1} for each depth's temperature time series---the results revealed statistically significant increasing warming trends. Shortly, we shuffled each of the time series under consideration, typically 1000 times, and calculated their slopes. The PDF of the slopes of the shuffled time series should be centered around zero where the spread of the PDF of the slopes may provide a measure for the uncertainty of the observed slope. 

\begin{figure}[t]
    \centering
    \includegraphics[width=\textwidth]{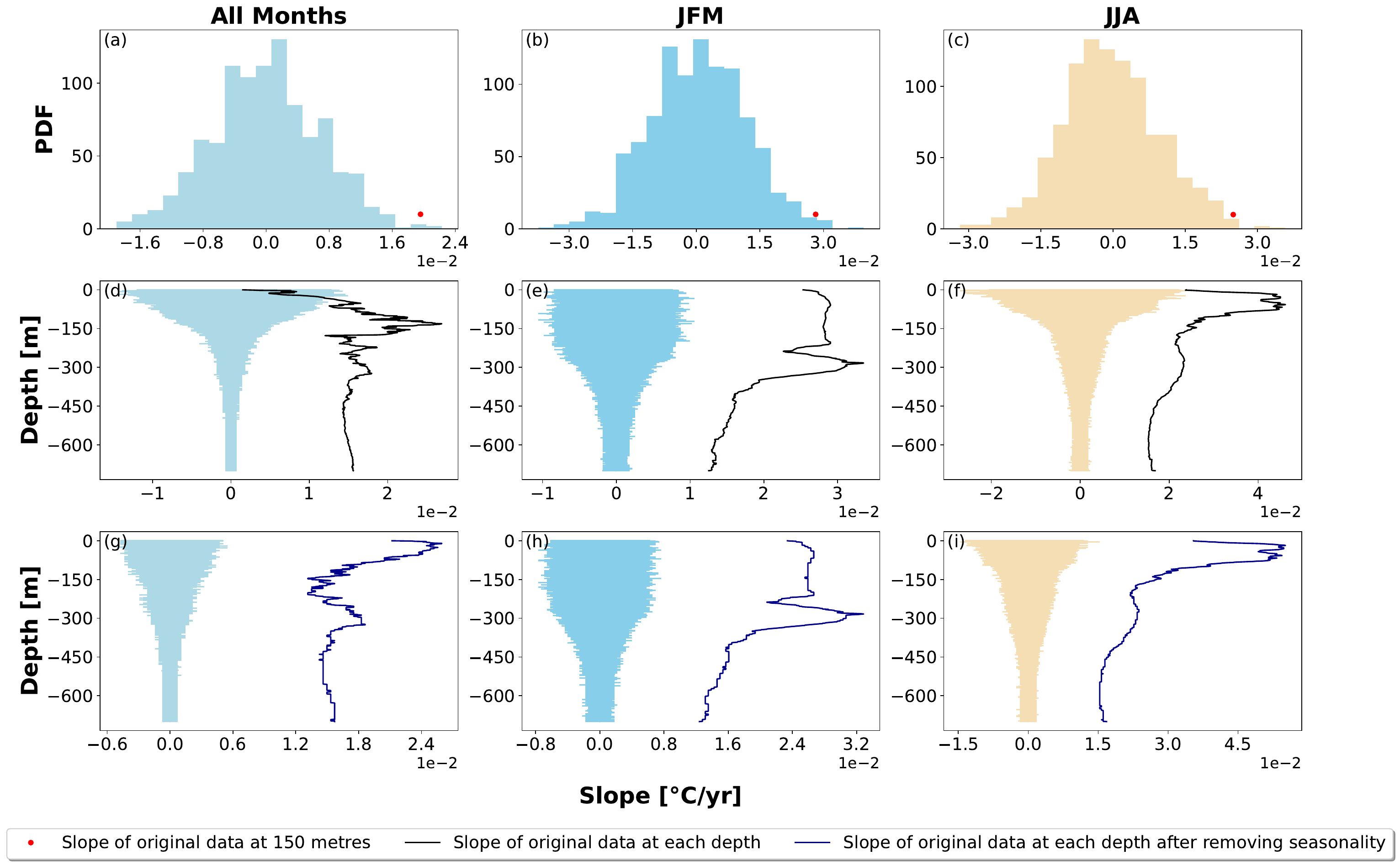}
    \caption{PDF of the slopes of the shuffled time series of 150 m depth for (a) the complete time series (shown in Fig.~\ref{fig:4}b), (b) of the winter months (January, February, and March stacked together), and (c) of the summer months (June, July, and August stacked together). The red dot in panels a-c indicates the slope of the original time series. The slope of the regression of the original time series versus depth (black line, in \degree{C}\,yr$^{-1}$) and the 25\%--75\% confidence interval (light blue shading) of the slopes of the shuffled time series versus depth based on the (d) original time series (e) the stacked data of the winter months (JFM) and, (f) the stacked data of the summer months (JJA). (g), (h), (i) Same as d-f but for the anomaly time series.}
    \label{fig:9}      
\end{figure}

Fig.~\ref{fig:9}a shows the PDF of the slopes of the shuffled time series of the depth of 150 m; the red point denotes the slope of the regression line of the original time series (shown in Fig.~\ref{fig:4}b). The slope of the original time series lays well outside the main part of the PDF of the slopes of the shuffled time series, indicating that the observed slope is significant. We have repeated the above procedure for all depths and present the results in Fig.~\ref{fig:9}d. Here, the slope of the original time series is indicated by the solid black line while the shading represents the 25\%--75\% confidence interval of the shuffled time series. The increasing trend in water temperature is highly significant except for the upper 50 m or so. 

Furthermore, to account for seasonal variations explicitly, we segregated the data into winter (January to March) and summer (June to August) months and performed the shuffling test to determine the significance of the trends (see Fig.~\ref{fig:9}b,c). As expected, the increasing warming slopes are higher across all depths when considering separately the winter and summer months and these slopes are highly significant (Fig.~\ref{fig:9}e,f). Analysis of each month separately yielded similar results (Supplementary Fig.~\ref{fig:12}). 

We repeated the above analysis on the anomaly time series of all the months and segregated summer and winter months, similar to~\ref{fig:9}d-f. In Fig.~\ref{fig:9}g-h the blue line indicates the slope of the regression of the original anomaly time series and the shaded region indicates the 25\%--75\% confidence interval of the PDF of the slopes of the shuffled anomaly time series. Almost all slopes of regression of the un-shuffled time series displayed in Fig.~\ref{fig:9} are positive and highly significant, confirming a consistent temperature rise across all depths.

In addition to the above shuffling test, we performed a ``weaker'' shuffling procedure in which the order of the time series within each year remains unchanged where only the order of the different years was chosen randomly (Supplementary Fig.~\ref{fig:13}). This shuffling procedure resulted in a significant water-warming trend. 
 
In summary, we find that the water column in the Gulf exhibits a significant warming trend. A trend analysis of the salinity time series resulted in much weaker trends in the upper part of the water column and significantly increasing trends in the deep ocean; see Supplementary Figs. \ref{fig:salinity1}, \ref{fig:salinity2}.

\subsection{Meteorological Data results}
\label{sec:3.2}

\begin{figure}[t]
    \centering
    \includegraphics[width=\textwidth]{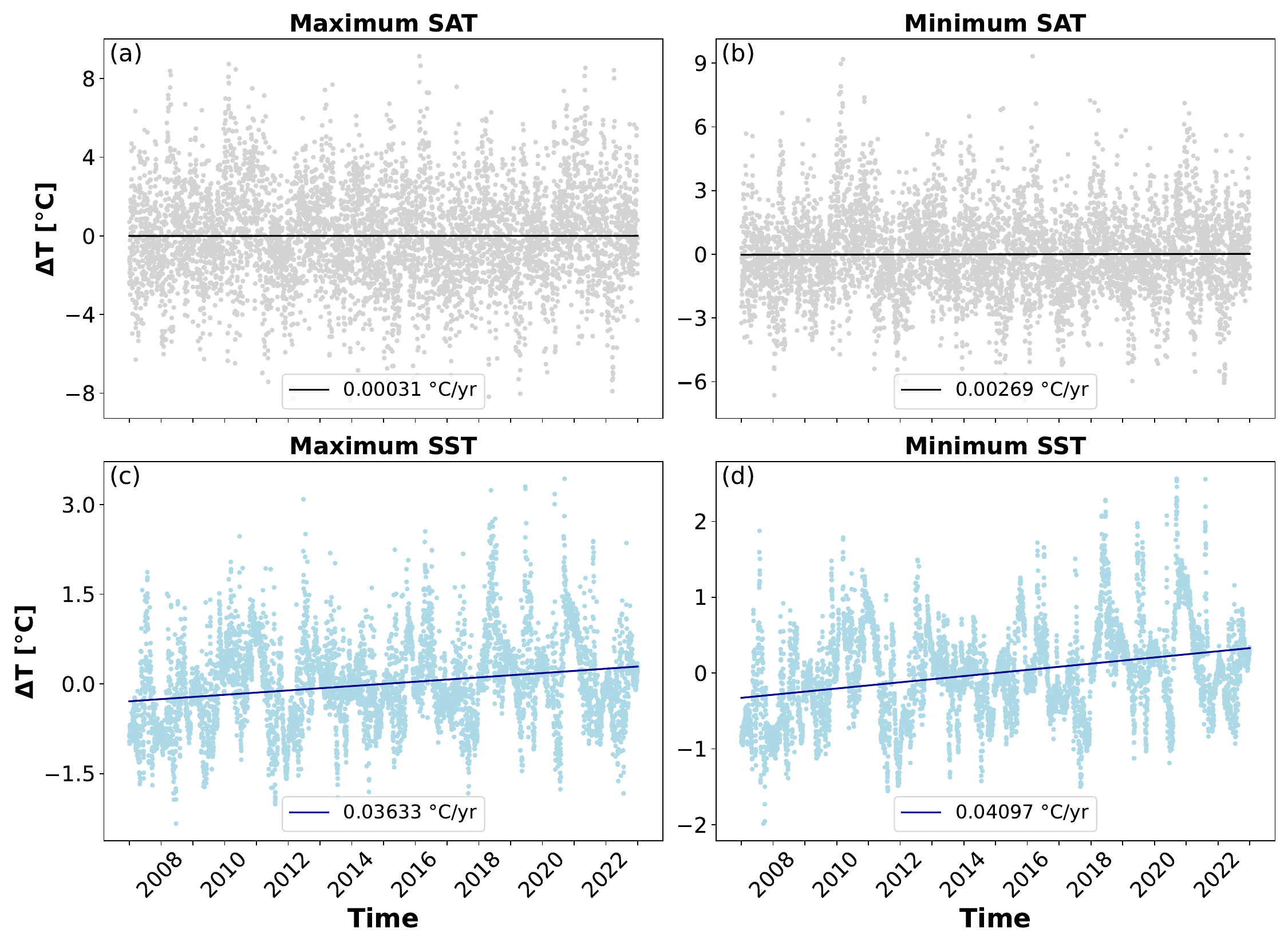}
    \caption{Anomaly time series of (a) daily maximum surface air temperature (SAT), (b) daily minimum SAT, (c) daily maximum SST, and (d) daily minimum SST. The corresponding original time series are shown in Fig.~\ref{fig:SAT-SST}.}
    \label{fig:6}       
\end{figure}

\begin{figure}[t]
    \centering
    \includegraphics[width=\textwidth]{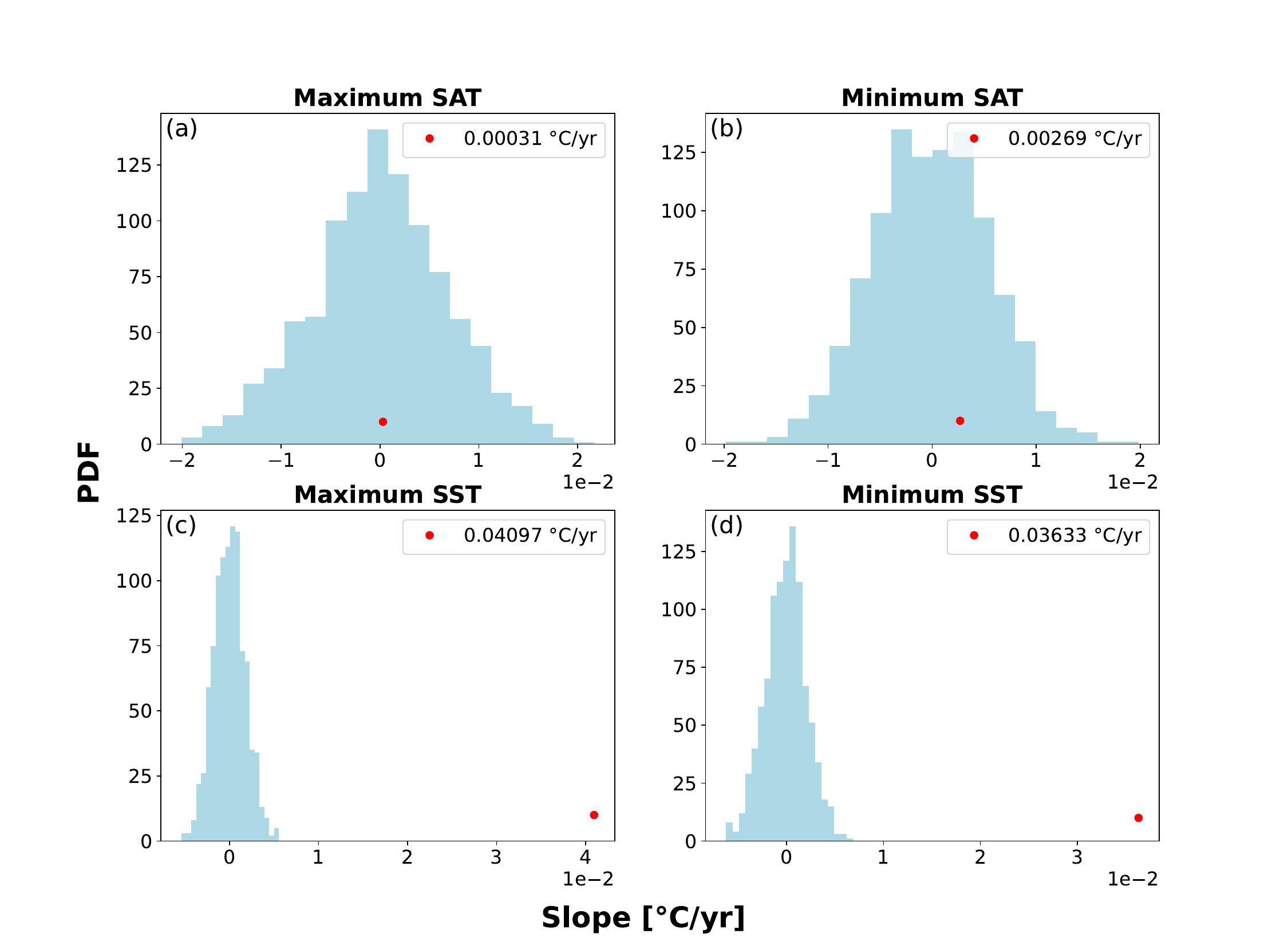}
    \caption{The PDF of the slopes of the shuffled time series of (a) daily maximum surface air temperature (SAT), (b) daily minimum SAT, (c) daily maximum SST, and (d) daily minimum SST. The red dot indicates the slope of a linear regression performed on the original time series. Note the significant trend for the SST time series and the absence of trends for the SAT time series.}
    \label{fig:6.1}       
\end{figure}

The meteorological data analysis is conducted using 10-minute averages of different variables, such as surface air temperature and SST. Additionally, we examined the daily (24-hour) maximum and minimum surface air temperature and SST. We also constructed the anomaly time series of the daily maximum and minimum time series of surface air temperature and SST (Fig.~\ref{fig:6}) and performed the trend analysis. We find that there is a significant increasing trend in the maximum and minimum SST (Fig. \ref{fig:6.1}c,d); similar to the above, the analysis was based on the slopes of linear regression of 1000 shuffled time series. The surface air temperature time series do not exhibit a similar increasing warming trend (Fig.~\ref{fig:6.1}a,b). The absence of trends in the surface air temperature series in contrast to the significant warming trends in the SST time series suggests that the local air temperature does not underlie the observed water warming.  In addition to this analysis, we conducted the ``weaker'' shuffling procedure in which the order of the time series within each year remains unchanged where only the order of the different years was chosen randomly (Supplementary Fig.~\ref{fig:16}) and found similar results of an increasing warming trend in the SST time series but not in the surface air temperature time series.

\begin{figure}[t]
    \centering
    \includegraphics[width=\textwidth]{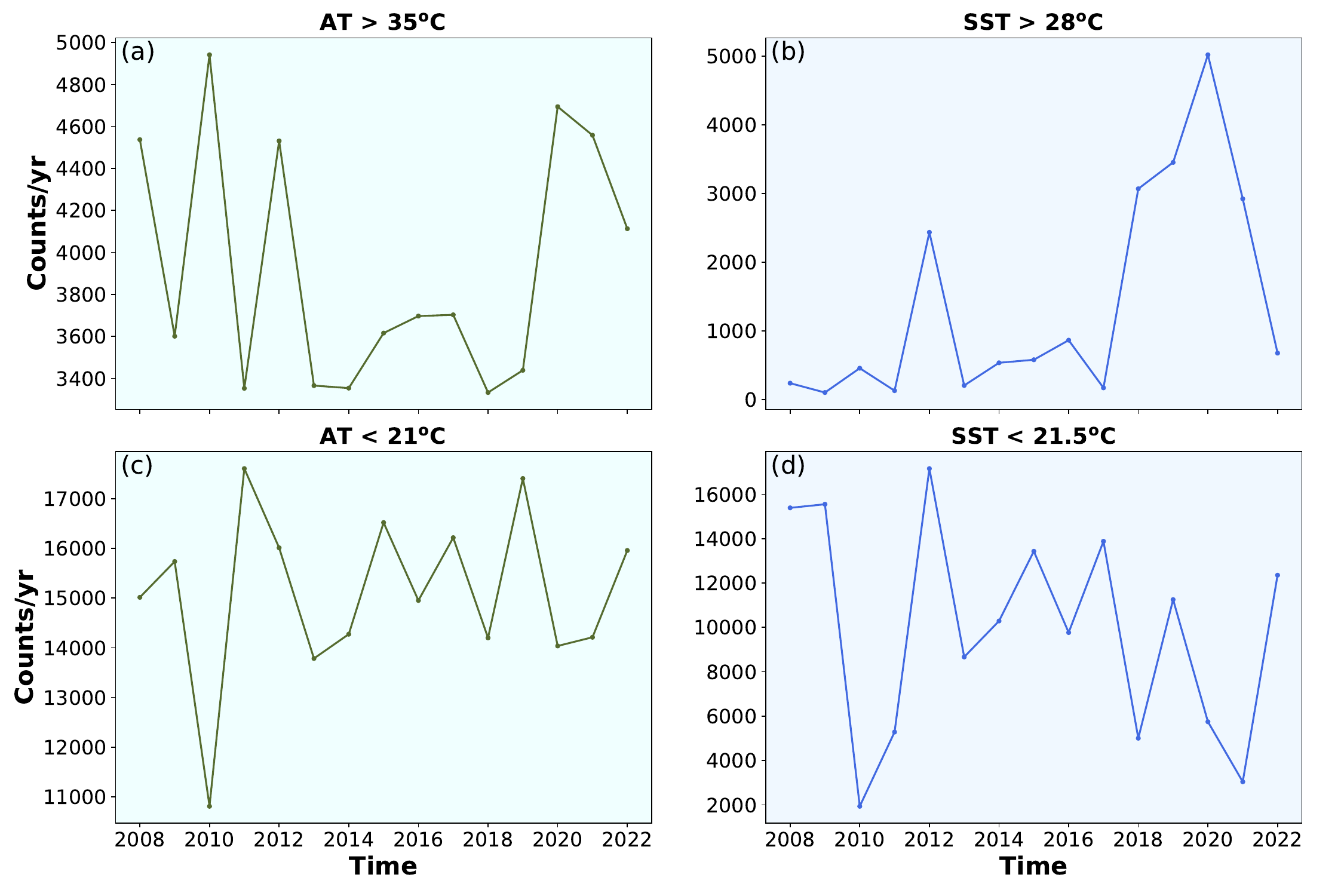}
    \caption{Counts (per year) of surface air temperature (a) above 35\degree{C} and (c) below 21\degree{C}. Counts (per year) of SST (b) above 28\degree{C} and (d) below 21.5\degree{C}. The counts are based on the 10-minute time series.}
    \label{fig:7}       
\end{figure}

Following the above, we adopted a different approach to studying possible temporal trends in surface air temperature and SST. The idea is to quantify the number of extreme cooling events and extreme warming events in the temperature time series; a reduction in extreme cooling events indicates a reduction in deep mixing events of the water column and hence warming of the entire water column. Based on the 10-minute data we counted the number of times the temperature dropped below a certain threshold and exceeded a certain temperature threshold. The results are presented in Fig.~\ref{fig:7} where it is apparent that while the analysis for the surface air temperature (Fig.~\ref{fig:7}a,c) does not yield noticeable trends, these of the SST exhibit a general increase in the number of warming events (Fig.~\ref{fig:7}b) and reduced number of cooling events (Fig.~\ref{fig:7}d). A noticeable exception is the warming event of 2010 that yielded a reduced number of cooling events both in the surface air temperature and SST (Fig.~\ref{fig:7}c,d). The above analysis is another indication of the warming of the Gulf's water. Yet, the absence of similar trends (or lack of correlations) in the surface air temperature counts (Fig.~\ref{fig:7}a,c) suggests that changes in the surface air temperature are not the primary cause for the warming of the water in the Gulf. 

In light of the findings detailed above, it is necessary to consider other factors that can influence the surface air temperature and SST differently. Local oceanographic processes, such as heat exchange with the atmosphere and coastal currents, can play a crucial role in modulating SST trends. Additionally, oceanic circulation may also lead to uncoupling between the local atmospheric and oceanic temperature variations. These are studied in more detail below.

Our findings are consistent with earlier research that has identified unique patterns for air temperature and SST in different geographical areas \cite{Decouple,Israel1970-22}. The decoupling of air temperature and SST trends can have implications for the local climate and marine ecosystems. Further investigations into the underlying mechanisms driving these trends are essential for gaining a comprehensive understanding of the climate dynamics in the Gulf area. To gain a deeper understanding of the process that causes the observed warming, we next examine the heat fluxes between the air and the sea.

\subsection{Heat Flux results}
\label{sec:3.3}

\begin{figure}[t]
    \centering
    \includegraphics[width=0.9\textwidth]{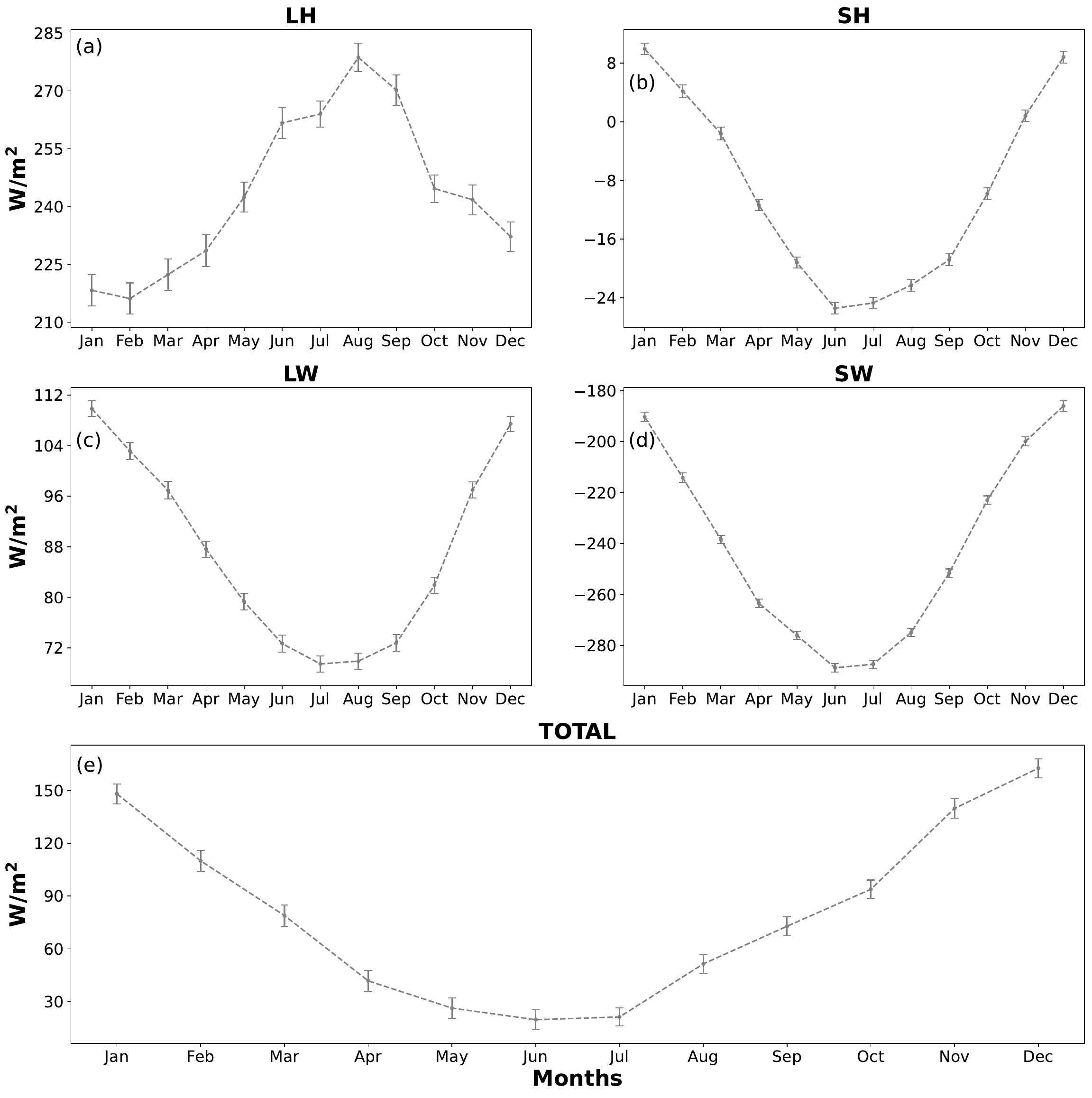}
    \caption{The seasonal cycle of the different heat fluxes. (a) Latent heat (LH), (b) Sensible heat (SH), (c) shortwave (SW) radiation, (d) longwave (LW) radiation, and (e) net heat flux radiation. The error bars signify the monthly standard deviation of the heat fluxes when averaged over all the years for each month.}
    \label{fig:8}        
\end{figure}

We calculated the different air-sea heat fluxes using the bulk formulas described in Sec.~\ref{1.3.3}; an exception is the shortwave radiation, which was directly measured. We constructed the climatological seasonal cycle of the latent heat (LH), sensible heat (SH), shortwave radiation (SW), and longwave (LW) radiation; see Fig.~\ref{fig:8} for the mean $\pm$ std seasonal cycle of the different fluxes. The LH is always positive as it is proportional to the difference between the saturated humidity at the sea surface to the actual humidity of the surface air---the saturated humidity is larger than the actual humidity. On the other hand, the SH is only positive during the winter months because it is constructed from the difference in SST and surface air temperature; during the winter the air temperature can be lower than the SST while the opposite occurs during the summer (see Fig.~\ref{fig:SAT-SST}). 

The results described above are in general agreement with the results of Ben-Sasson et al. \cite{NathanPaldor} who used one-year data of ocean temperature up to a depth of 750\,m as well as meteorological data to analyze the Gulf's heat balance using specific bulk formulas, revealing insights into the heat flux and evaporation processes in the Gulf. Yet, our analysis is based on 16 years of data while the analysis of Ben-Sasson et al. \cite{NathanPaldor} was based on one year of data. The net heat flux shown in Fig.~\ref{fig:8}e is positive throughout the year such the that annual mean net heat flux is not zero as one would expect. This finding is similar to \cite{NathanPaldor} and suggests that warm water masses originating from the southern parts of the Gulf and the Red Sea contribute to the observed temperature changes in the Gulf and close the energy budget. Fig.~\ref{fig:8} provides further insights into the seasonal variations in the average latent heat. Interestingly, we observed a distinct peak in LH over the summer months, in contrast to Ben Sasson et al. \cite{NathanPaldor}; this discrepancy may originated in the much shorter time series analyzed by Ben Sasson et al. \cite{NathanPaldor}.

\section{Summary and Conclusion}
\label{sec:summary}

The multi-depth analysis of ocean temperature in our study has revealed a prominent and consistent warming in the past 20 years or so, resulting in less frequent deep mixing events. This is supported by the temperature anomaly time series that resulted in a larger warming trend in the upper water layers compared to deeper layers. This temperature increase is also evident in the SST time series and other measures but is absent in the surface air temperature. This finding suggests that the local atmospheric forcing may not be the primary driver underlying the water warming in the Gulf. We note that surface fluxes are also not the driver of summer warming and stratification. We also calculated the heat flux budget in the Gulf and found that the net heat flux is not zero, supporting the idea that the observed water warming is due to water advection from southern latitudes. The reported warming may have significant implications for the local marine ecosystem and can contribute insights to the broader understanding of the effect of remote climate change on the local marine environment.

The hypothesis of horizontal advection aligns with prior research on the Red Sea \cite{cantin,Berman2022,Wolf-Vecht:1992}. These studies reported an increase in SST in tropical regions, ranging from around 0.014 \degree{C} yr$^{-1}$ to 0.034 \degree{C} yr$^{-1}$ since the mid-1970s, which has resulted in more frequent and severe coral bleaching and mortality, thereby raising concerns about the survival of coral reefs. More specifically, Wolf-Vecht et al. \cite{Wolf-Vecht:1992} developed a one-dimensional convective model to study the structure of the water column in the Gulf. The model accurately replicates the thermal structure in the Gulf with a slight temporal lag in the summer thermocline; still, it failed to reproduce the observed salinity minimum. Introducing 40.3\% advected Red Sea water rectifies the salinity discrepancy and may explain the early thermocline development due to the influx of warmer water. Another study by Cantin et al. \cite{cantin}, investigated the SST in the central Red Sea and its correlation with reef-building coral species, Diploastrea heliopora, employing three-dimensional computed tomography. From the analysis of their data combined with the Intergovernmental Panel on Climate Change (IPCC) climate model simulation, they observed a significant temperature gradient between the Gulf and the central Red Sea. They also predict that with the current warming trend, this coral could cease growing altogether by 2070. 
Furthermore, our research findings are in harmony with the observations and conclusions presented in the IPCC report on the Red Sea, which also indicates a higher average SST in the central Red Sea compared to the Gulf.

In conclusion, our comprehensive analysis of ocean temperature data in the Gulf highlights a significant rise in temperature over the entire water column. The absence of a direct correlation with local atmospheric conditions and the observation that the local net heat flux is not zero suggest that other factors, such as horizontal advection, are likely influencing the observed water warming. Our study contributes to the growing body of knowledge regarding oceanic processes and their implications for regional and global climate dynamics. Understanding these mechanisms is critical for addressing the challenges of climate change and its potential impacts on marine ecosystems and coastal communities.

\newpage


\section*{Acknowledgements}
HG was supported by a grant from the Ministry of Science and Technology. We thank Asaph Rivlin, Modi Pilersdorf, the Israel National Monitoring Program at the Gulf of Eilat, and The Interuniversity Institute for Marine Sciences in Eilat for access to infrastructure and services.

\section*{Declarations}

\subsection*{\bf Ethics approval and consent to participate}

Not Applicable.

\subsection*{\bf Consent for publication}

Not Applicable.

\subsection*{\bf Competing Interests}

The authors have no relevant financial or non-financial interests to disclose.

\subsection*{\bf Funding}

Supported by a grant from the Ministry of Science and Technology

\subsection*{\bf Availability of data and materials}

CTD and Meteorological data analyzed in this research work can be obtained from the website of The Israel National Monitoring Program at the Gulf of Eilat. 

CTD Data - \url{https://www.meteo-tech.co.il/EilatYam_data/ey_ctd_data_download.asp}

Meteorological Data - \url{https://www.meteo-tech.co.il/eilat-yam/eilat_en.asp}


\bibliography{sn-bibliography}

\setcounter{figure}{0}
\renewcommand{\thefigure}{S\arabic{figure}}

\begin{figure}[ht]
    \centering
    \advance\leftskip-0.0cm
    \includegraphics[width=\textwidth]{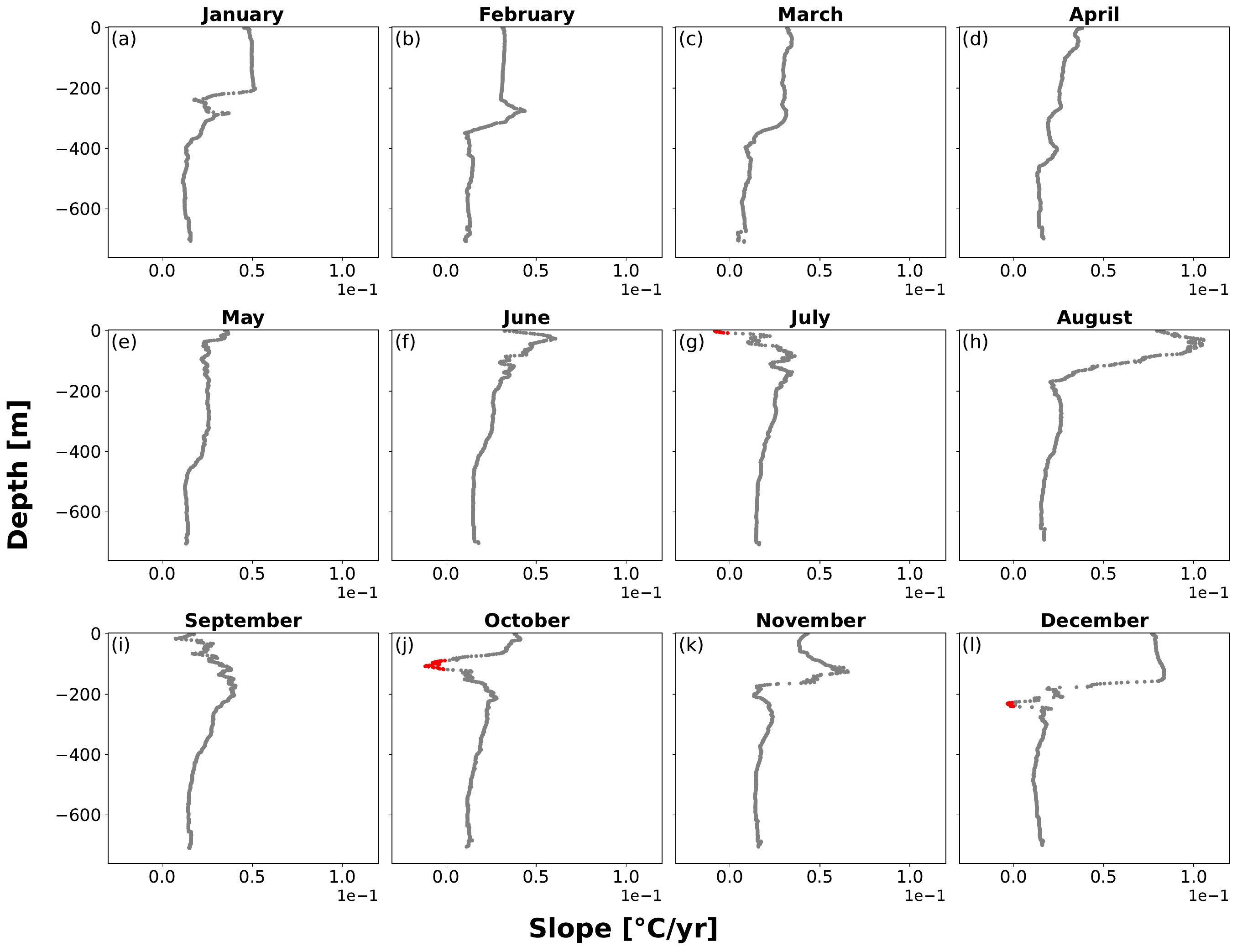}
\caption{Slopes of linear regression of the water temperature time series after month-wise segregation. The red portion in the curves represents negative slopes, which are present only in certain depths of July, October, and December. 
}
\label{fig:12}     
\end{figure}

\begin{figure}[ht]
    \centering
    \advance\leftskip-0.0cm
    \includegraphics[width=\textwidth]{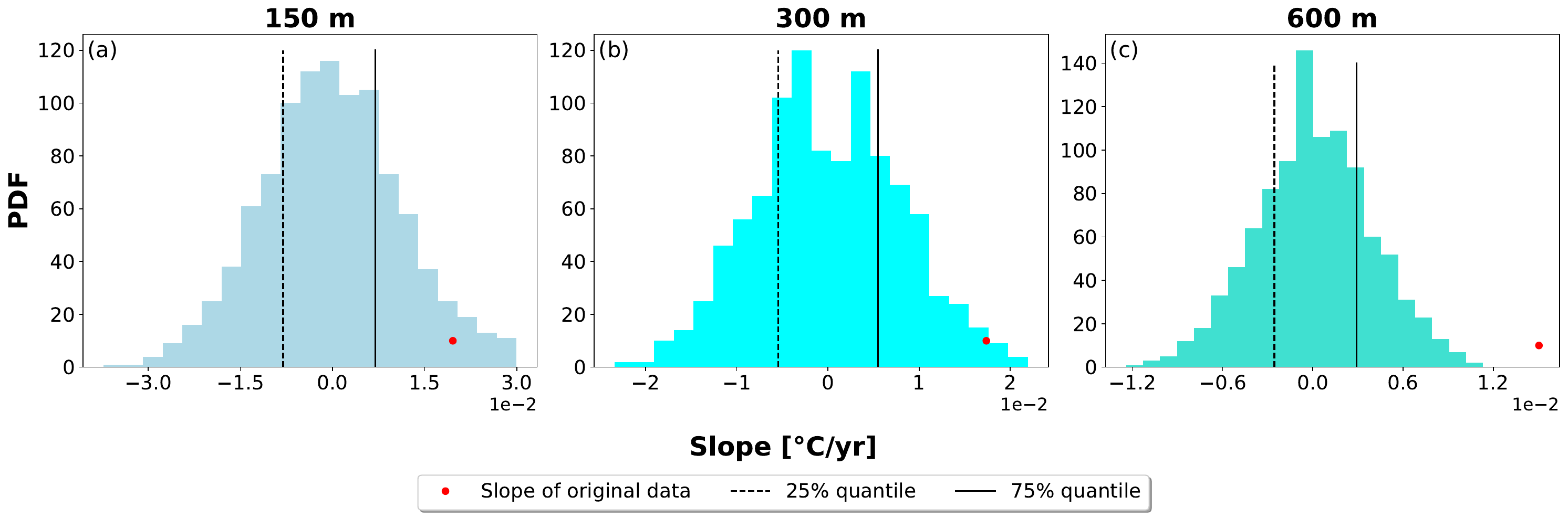}
\caption{The results of shuffling the water temperature time series by years; i.e., the time series of each year's water temperature remains unchanged while the order of different years is randomized. The PDF of the slopes of these shuffled time series at a depth of (a) 150 m, (b) 300 m, and (c) 600 m. The dashed and solid lines signify the 25\% and 75\% quantiles and the red dot denotes the slope of the original time series, indicating a significantly increasing warming trend even when using this weak shuffling.}
\label{fig:13}     
\end{figure}

\begin{figure}[ht]
    \centering
    \advance\leftskip-0.0cm
    \includegraphics[width=\textwidth]{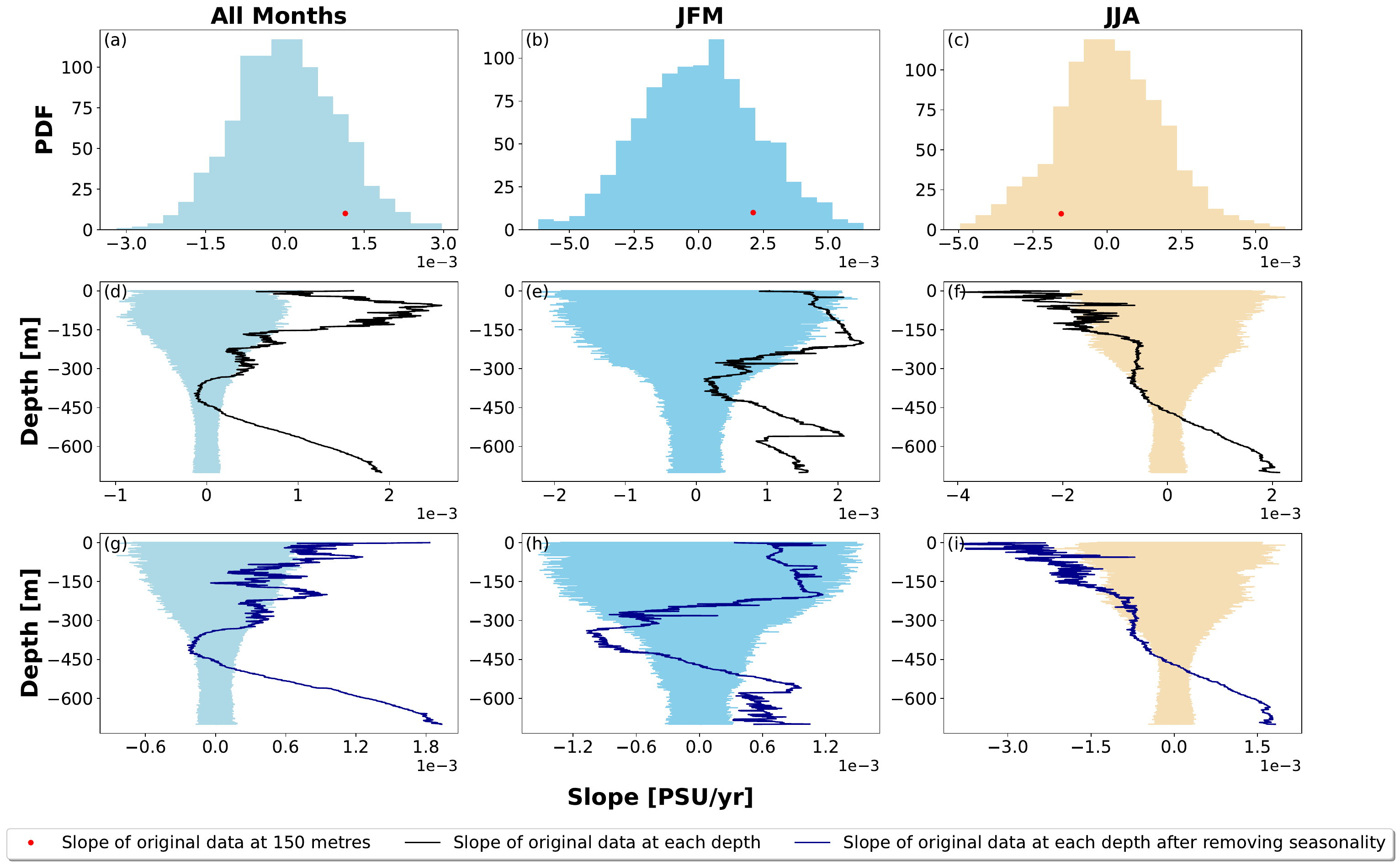}
\caption{Same as Fig. \ref{fig:9} for Salinity. Probability distribution function (PDF) of the slopes of the shuffled time series of 150 m depth for (a) the complete time series, (b) of the winter months (January, February, and March stacked together), and (c) of the summer months (June, July and, August stacked together). The red dot in panels a-c indicates the slope of the original time series. The slope of the regression of the original time series versus depth (black line, in PSU yr$^{-1}$) and the 25\%--75\% confidence interval (shading) of the slopes of the shuffled time series versus depth based on the (d) original time series (e) the stacked data of the winter months (JFM) and, (f) the stacked data of the summer months (JJA). (g), (h), (i) Same as d-f but for the anomaly time series.}
\label{fig:salinity1}     
\end{figure}

\begin{figure}[ht]
    \centering
    \advance\leftskip-0.0cm
    \includegraphics[width=\textwidth]{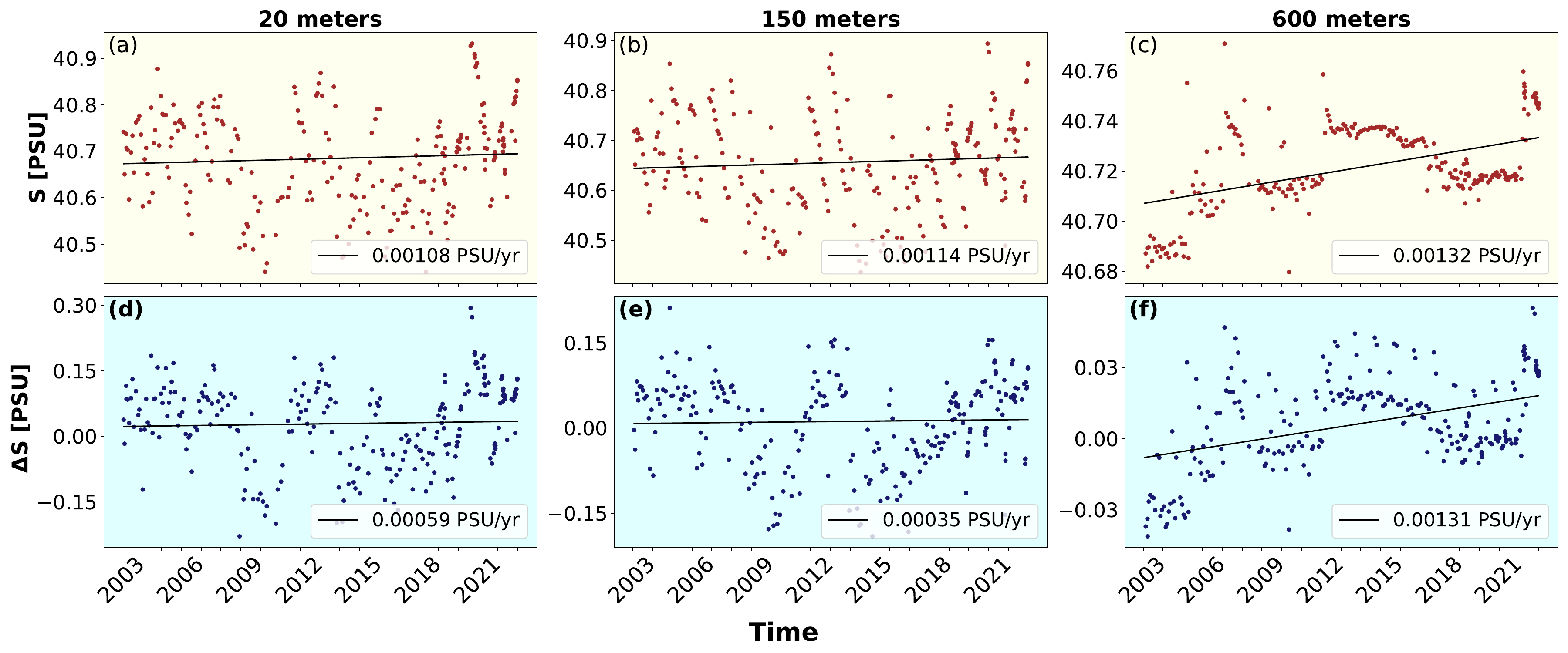}
\caption{Same as Fig. \ref{fig:4} for the salinity.}
\label{fig:salinity2}     
\end{figure}

\begin{figure}[ht]
    \centering
    \advance\leftskip-0.0cm
    \includegraphics[width=\textwidth]{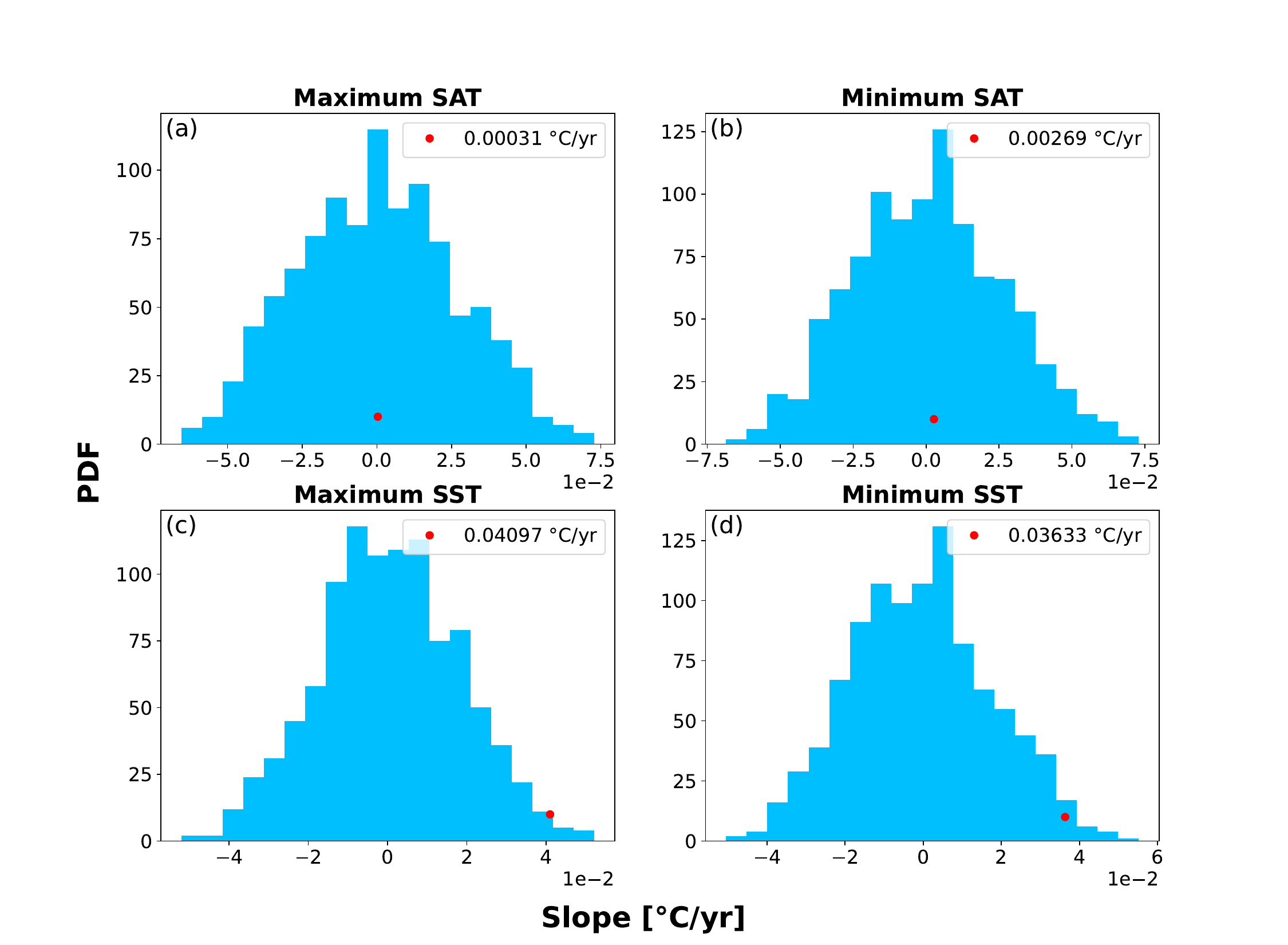}
\caption{Same as Fig. \ref{fig:13} for the (a) daily maximum surface air temperature (SAT), (b) daily minimum SAT, (c) daily maximum SST, and (d) daily minimum SST. Note that the analysis indicates a non-significant trend for the SAT and a significant trend for the SST.}
\label{fig:16}     
\end{figure}

\begin{figure}[ht]
    \centering
    \advance\leftskip-0.0cm
    \includegraphics[width=\textwidth]{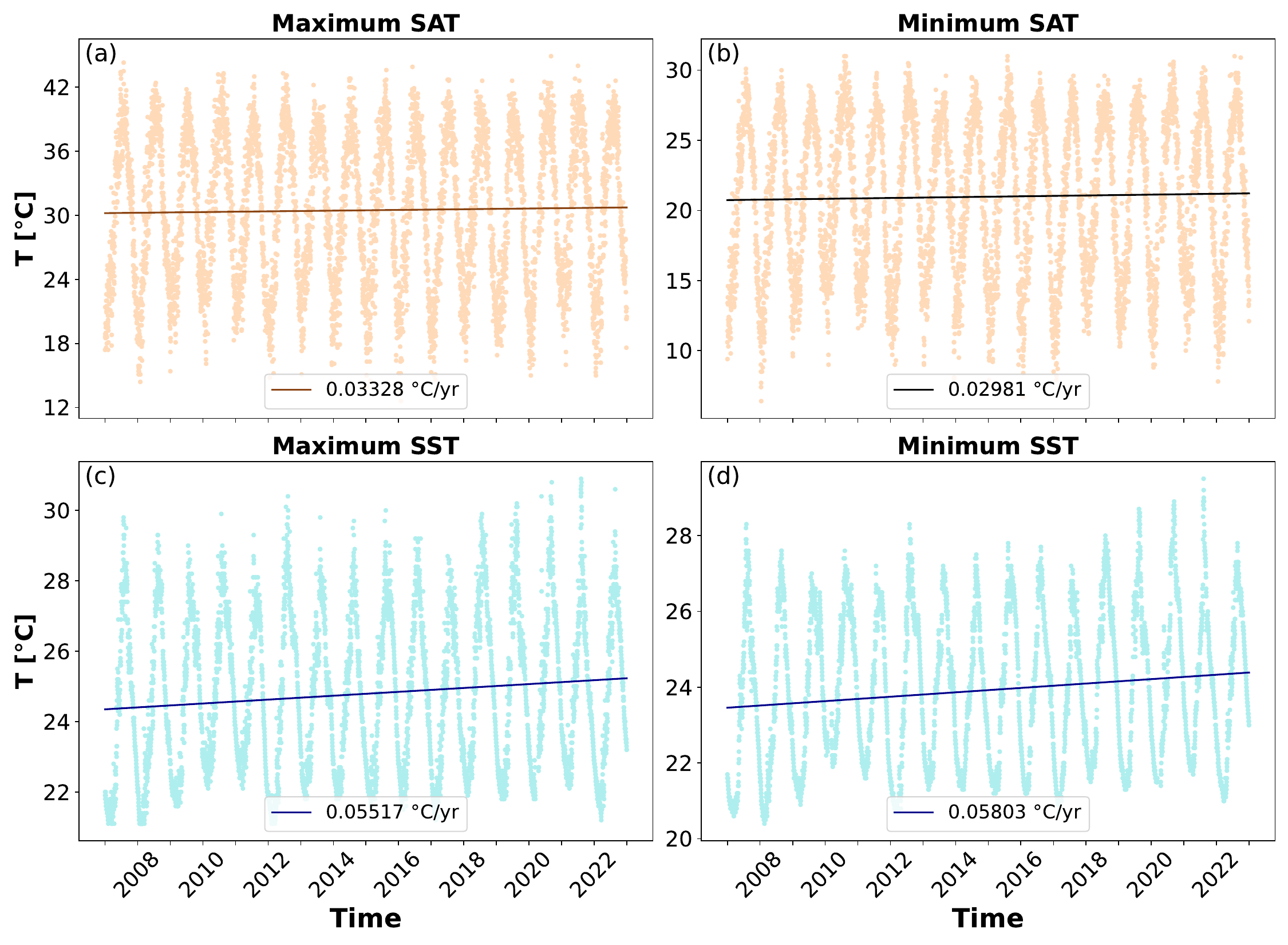}
\caption{Same as Fig. \ref{fig:6} for original data of maximum and minimum of Surface Air Temperature (SAT) and Sea Surface Temperature (SST).}
\label{fig:SAT-SST}     
\end{figure}

\end{document}